\documentclass{article}
\usepackage{graphicx}

\newcommand{\fig}[1]{Fig.~\ref{fig.#1}}

\newcommand{\tbl}[1]{Table~\ref{table.#1}}

\newcommand{\figlabel}[1]{\label{fig.#1}}
\newcommand{\tbllabel}[1]{\label{table.#1}}

\newcommand{\ket}[1]{ {\left| #1 \right\rangle} }

\title{Quantum Solution of Coordination Problems}

\author{Bernardo A. Huberman\thanks{email: huberman@hpl.hp.com} and Tad Hogg\\HP Labs, Palo Alto, CA 94304}

\begin{document}
\maketitle

\begin{abstract}
We present a quantum solution to coordination problems that can be
implemented with existing technologies. It provides an alternative
to existing approaches, which rely on explicit communication,
prior commitment or trusted third parties. This quantum mechanism
applies to a variety of scenarios for which existing approaches
are not feasible.

Keywords; coordination games, mechanism design, quantum
information

Classification C71,C72
\end{abstract}

\newpage
The existence of multiple equilibria in economic systems can lead
to coordination failures and consequently to inefficient outcomes.
Examples that have been extensively studied range from firms
having to decide whether or not to enter a competitive market and
and how to position their offerings, to the coordinated resolution
of social dilemmas involved in the provision of public goods.

Coordination problems have long been studied in the context of
game theory~\cite{schelling60,fudenberg00,camerer03}, where the
coordination game is specified by a payoff matrix which yields
several Nash equilibria. These equilibria can at times give the
same payoff to all players, in which case the problem is for them
to agree on which one to coordinate, or different payoffs, leading
to a competitive coordination game in which players prefer
different equilibria.

A simple example of a cooperative coordination game is that of two
people having to choose driving on the left or the right side of
the road, for which the payoff matrix is shown in \tbl{payoffs}.
As can be seen, there are two Nash equilibria, with equal payoffs,
corresponding to both drivers choosing the same side of the road.
The coordination problem consists in both drivers finding a way to
agree on which side of the road to drive.

\begin{table}[h]
\begin{center}
\begin{tabular}{c|cc}
choice & L  & R \\ \hline
L   & $2,2$    & $-3,-3$    \\
R   & $-3,-3$    & $2,2$\\

\end{tabular}
\end{center}
\caption{\tbllabel{payoffs}Payoff structure for two driving
choices: left (L), right (R). Each row and column corresponds to
choices made by the first and second players, respectively, and
their corresponding payoffs. }
\end{table}

This and many other instances of coordination problems can be
solved in several ways. A first solution resorts to a trusted
third party who knows the preferences of the participants and is
given the authority to pick an equilibrium which is then
broadcasted to the players. In the case of competitive
coordination problems the trusted third party may also have
enforcement powers, since some players may wish to move the group
to a more preferable equilibrium.

Another solution to coordination problems involves communication
among players so that they can negotiate a choice. In the case of
cooperative games even one player flipping a coin and broadcasting
the result as the corresponding choice provides an effective
solution. In a competitive setting, the negotiation might be such
that the players wish to choose their equilibria at random as it
would then be perceived as a fair choice. This case would require
a trusted mechanism of coin flipping over a communication line,
which can be enforced through cryptographic protocols.

A third mechanism for solving coordination problems invokes social
norms, in which common knowledge of the participants' preferences
can distinguish a given equilibrium from the others, as in the
case of choosing the largest river as a boundary between two
countries. Such distinguished equilibria are often called focal
points~\cite{schelling60,huyck97}.

While these mechanisms can solve in principle coordination
problems, there are times when none of these options are
available, either because they are too expensive, slow or
difficult to implement, or because privacy worries prevent the
participants from using any of these options. Even worse, a
constraint from a larger context, such as the need to use a mixed
strategy, might make it disadvantageous for players to have their
choices revealed in advance. Furthermore in cases where
communication between parties is asynchronous there is the
additional problem of achieving common
knowledge~\cite{fudenberg00}, i.e. all parties know that the
others know how to act. A simple example is that of using email to
coordinate a meeting when one is not sure that recipients have
read their emails and acknowledgements before the start of the
meeting~\cite{rubinstein89,binmore00}.

It would appear that in these instances the only choice left for
the participants is to choose at random which strategy to pursue,
which would lead to many instances of coordination failures and a
consequent reduction in their respective payoffs. Nevertheless, as
we now show, there is an alternative and superior solution, which
resorts to quantum mechanics to solve coordination problems
without communication, trusted third parties or prearranged
strategies. Moreover, this quantum solution is implementable in
the real world, thus making coordination problems easier to solve.

The way quantum dynamics allows for the practical solution of a
coordination problem is via the generation of particles in
entangled states. Quantum entanglement results in the appearance
of specific quantum correlations between parts of a composite
system, which can be exploited for quantum information
processing~\cite{chuang00}. In particular, parametric-down
conversion techniques can produce twin photons which are perfectly
quantum correlated in time, space and often in
polarization~\cite{kwiat00}. These photons can then be physically
separated at arbitrary distances so that each participant gets one
of the entangled particles.

In the simplest case, where players face two choices, {0}, and
{1}, they can use entangled particles with two physically
observable states, such as their spin or polarization. As shown in
\fig{entanglement}, the corresponding state is then given by a
superposition of the two correlated possibilities, denoted as
$(\ket{0,0}+\ket{1,1})/\sqrt{2}$. At a time of their choosing,
each participant observes the state of their particle, resulting
into a {0} or a {1} state, and makes the corresponding choice. The
key aspect that makes this technique different from random choices
is that entanglement implies a definite correlation between the
two measurements, i.e. both players get either a $0$ or a $1$,
irrespective of the spatial separation between them, and without
communication.

\begin{figure}
\begin{center}
\includegraphics[scale=0.5]{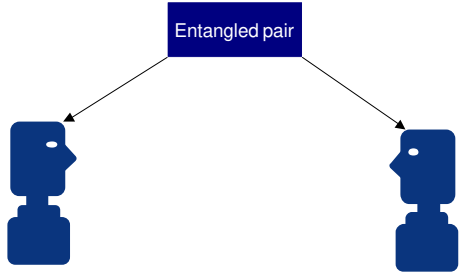}
\end{center}
\caption{\figlabel{entanglement} A source of entangled photons
sends one each to the participants of a coordination game.}
\end{figure}

One could argue that this correlation could also be achieved
classically by flipping two coins in advance, hiding them into two
boxes given go the players  which they then open at some later
time. However, unlike the entangled quantum solution we just
described, this procedure predetermines the outcome, which may not
be desirable if the players wish to defer the choices as long as
possible. In this case an adversary might learn how the players
will choose long before they actually do, and thus adjust its
strategy accordingly.

To illustrate this consider the case of two players trying to
coordinate on a mixed strategy against a third one without
resorting to previous agreement or communication, as in the case
of a coordinated attack on a rival or enemy. For the sake of
example, consider the game of rock, paper, scissors, in which the
two allied players must make the same choice to have any chance of
winning. If the allies make different choices their payoffs are
zero and the third player gets a payoff of 1. When the two allies
make the same choice the payoff to the allies and the third player
are given the payoff matrix of the rock, paper scissors game,
which is shown in \tbl{mixedpayoffs}.

\begin{table}[ht]
\begin{center}
\begin{tabular}{c|ccc}
choice & rock  & paper & scissors\\ \hline
rock   & $0,0$    & $0,1$ & $1,0$   \\
paper   & $1,0$ &$0,0$ &$0,1$\\
scissors   & $0,1$    & $1,0$  &$0,0$\\

\end{tabular}
\end{center}
\caption{\tbllabel{mixedpayoffs}Payoff structure for the rock,
paper, scissors for the pair of allied players against the third
player. Each row and column corresponds to choices made by the
pair (assuming that they are the same) and their opponent,
respectively, and their corresponding payoffs. For example, the
entry of the first column, second row corresponds to the allies
both choosing paper and the third player choosing rock.}
\end{table}

This game has the feature that no single choice is best, i.e.
there is no pure strategy Nash equilibrium. Instead, the best
strategy for rational players is to make the choices randomly and
with equal probability, which gives it a mixed strategy Nash
equilibrium with expected payoff of 1/3.

For the full game without coordination the pair of allied players
only has 1/3 chance of making the same choice, and another 1/3 to
win against their opponent, leading to an expected payoff of 1/9.
If they can be perfectly coordinated their payoff would be 1/3. In
this example it is necessary to play random choices because any a
priori commitment between the allied pair would no longer be a
random strategy, and therefore discoverable by observation. If
they instead use a pseudorandom number generator with a common
seed, it could be compromised by the opponent discovering either
the random pattern or the seed. On the other hand, if one of the
allied pairs were to use a perfect private coin toss and
communicated it to the other, it would risk being detected or
jammed, leading to loss.

Using the quantum mechanism the players both can have undetectable
randomness in their choices, no communication between them and
still maintain complete correlation at every period of the game.
Entanglement thus offers a way for the players to get correlated
random bits they can use in addition to any public, broadcast
information, without communication or prior agreement.

Furthermore, this quantum solution cannot be achieved via a
classical simulation since we are requiring the absence of any
communication among the participants. This is unlike the situation
with other quantum games proposals~\cite{vanenk02}.

Thus, quantum information processing, which already offers the
potential for improved computation, cryptography and economic
mechanisms~\cite{eisert98,meyer99,du02,benjamin01,lee02a} can lead
to a perfect solution of complex coordination problems without
resorting to the complex signaling procedures that have been
discussed since they were first studied systematically. In
particular it solves the problem of achieving common knowledge in
the presence of asynchronous communication.

This quantum solution of a coordination problem is not just a
theoretical construct, as it can be implemented over relatively
large distances. It has been recently shown that it is possible to
produce pairs of entangled photons using parametric down
conversion, that can be sent separately over distances over many
kilometers. If the lifetime of the entangled state is long, each
participant can then receive an entangled photon and perform a
polarization measurement later, thus not having to communicate
with each other during the whole procedure. On the other hand, if
the lifetime of the entangled state is shorter than the period of
the game, photons can be regenerated periodically, thereby
requiring a transmission channel from the source to the
participants (but not between the participants). In this case the
advantage lies not in avoiding the possibility of blocked
communication by an adversary, but in avoiding the detection of a
coordinated solution. This makes for a feasible quantum solution
to coordination problems that can be implemented with current
technology, in contrast with most schemes for arbitrary quantum
computation.

This scheme can also be extended to situations involving many
participants. If the problem can be decomposed into independent
pairwise coordination that can be then coordinated at a higher
level (hierarchy) then the solution we described above can be
applied to each pair. More interestingly, it has been recently
shown~\cite{sackett00} that it is possible to create entangled
states of many particles in a single step and on demand, which
implies that coordination problems involving many participants can
also be solved using the scheme proposed in this paper.

Finally, this quantum approach to coordination games is more
general than it may first appear, as it can also be applied to a
variety of economic situations that involve achieving some or
partial coordination among members of a group.

One example is several groups participating in an auction in which
the value of an item to a person depends on what others in their
group get. For example imagine bidding for construction tools that
members of a group share and the auction is for each item
separately. In this case, the valuation depends on the
complementarity of the goods that the whole group gets, rather
than who in the group gets each item, While more complicated than
the pure coordination game we discussed because, this problem also
involves a bidding strategy, and thus the need for the group to
coordinate without signaling to other groups. The coordination
part of this problem can be solved by having a source produce a
quantum state $\ket{S}$ given by an entangled state that for the
case of two particles could be written as
\begin{equation}
\ket{S}= a \ket{A A} + a \ket{B B}+ b \ket{A B} + b \ket{B A}
\end{equation}
with the constants, $a, b$, chosen to favor a particular outcome
and subject to the normalization condition $2|a|^2 + 2|b|^2 = 1$.
Notice that these coefficients allow balancing the two parts of
the utilities involved in this game, the desire for coordination
and for low cost to the participants. For example, as cost
difference increases, one could reduce $a$, and increase $b$. In
other words, given a cost difference between items, one can pick a
suitable superposition of states.

Another situation where our quantum mechanism could be useful is a
coordination problem in which each player does not know others
involved in the game, or they wish to remain mutually anonymous
and avoid communication. If the interested players are known to be
members of a larger group~\cite{adar01}), and entangled states are
easily distributed among members of the larger group, those
players interested in coordinating their activities can use the
entangled states to ensure all players make the same choice.

As we have shown, the utilization of simple properties of quantum
states gives a solution to coordination problems that does not
require communication, trusted third parties involved in the
decision making or prior commitment. Equally interesting, this
solution is achievable with today's technology and opens the
practical use of quantum entanglement in real world problems.

\end{document}